\documentclass[twocolumn,prb,superscriptaddress]{revtex4}
\usepackage{graphicx}
\usepackage{dcolumn}
\usepackage{bm}
\usepackage{amsmath}
\usepackage{times}
\usepackage{color}
\usepackage[breaklinks=true,colorlinks,citecolor=blue,linkcolor=blue,urlcolor=blue]{hyperref}

\makeatletter

\newcommand{\Rmnum}[1]{\expandafter\@slowromancap\romannumeral #1@}
\makeatother

\begin{document}

\title{Identification of Griffiths-like phase and its evolution in Cr substituted pyrochlore iridates $Y_2Ir_2O_7$}
\author{Vinod Kumar Dwivedi}
\email{vinodd.iitbombay@gmail.com}
\affiliation{Materials Science Programme, Indian Institute of Technology Kanpur, Kanpur 208016, India}
\affiliation{Department of Physics, Indian Institute of Science, Bengaluru 560012, India}



\begin{abstract}

We report the Griffiths phase (GP)-like state along with cluster-glass-like state in geometrically frustrated antiferromagnetic $Cr$ substituted $Y_2Ir_2O_7$ pyrochlore iridates. The strength of GP-like behaviour increases with substitution. Interestingly, isothermal remanent magnetization suggests the Ising-like interaction of spins in GP region. The GP-like state is not attributed to the structural disorder as substitution of Cr does not induces any structural change. Then the spin coupling between $Cr^{3+} \leftrightarrow Cr^{3+}$, $Ir^{4+} \leftrightarrow Ir^{4+}$, $Ir^{4+} \leftrightarrow Cr^{3+}$ and $Ir^{4+} \leftrightarrow Ir^{5+}$ leads the competition between antiferromagnetic and ferromagnetic correlations. It give rise to Ruderman-Kittel-Kasuya-Yosida (RKKY)-like interaction between $Cr^{3+}$ local magnetic moments mediated by itinerant $Ir$ conduction electrons, hence, as a result GP and cluster-glass-like states emerges. 
\end{abstract}

\maketitle
\section{Introduction}
A spin glass (SG) is known as a randomly distributed mixed interacting bonds (antiferro and ferromagnetic) characterized  by a collective freezing of the spins at a definite temperature $T_{SG}$ below which a highly irreversible metastable frozen state appears without the usual magnetic long range ordering, where each and every spin reside in a frustrated state~\cite{Binder}. Spin glasses show a phase transition from a high temperature paramagnetic (PM) state into low temperature glassy state with exponential relaxation of spins. Long back, it have been reported that low temperature magnetic state fuses into a state whose correlation functions follow non-exponential tails~\cite{Mohit} due to the emergence of same un-frustrated clusters favour to the growth of Griffiths phase~\cite{Bray,Griffiths} in the magnetization, sandwiched between the spin-glass and paramagnetic state.

The Giriffiths phase (GP) was initially suggested for randomly diluted Ising ferromagnets, where only a tiny part of the lattice sites are filled with spins and remaining fractions are either empty or occupied with non-magnetic ions. Further, it has been shown that disorder suppresses the magnetic transition from its clean value of $T^\ast$ (Griffiths temperature) to long-range magnetic ordering temperautr $T_C$~\cite{Bray}.  GP is characterized by formation of magnetically ordered rare regions within the global paramagnetic (PM) matrix at $T_C < T < T^\ast$. Such a system contains sharp downturn at high temperature regime in inverse magnetic susceptibility 1/$\chi$ vs T curve below $T^\ast$~\cite{Deisenhofer}. The physics of GP is closely related to quenched disorder and competing interactions~\cite{Deisenhofer,Salamon,Pramanik,Vinay,Vinod1,Magen,Ouyang}. In fact, GP behaviour have mostly been reported in the diluted ferromagnetic systems with positive value of Curie-Weiss (CW) temperature $\theta_{CW}$, however, there are {\it{very limited}} experimental reports on GP exhibiting negative value of CW temperature $\theta_{CW}$ in antiferromagnetic (AFM) systems~\cite{Ouyang,Jitender,Ghosh,Sampathkumaran,Arkadeb}.

In pyrochlore iridates $R_2Ir_2O_7$ (R = Y, Bi, Rare earth elements), the interplay of spin-orbit coupling, electronic correlation and crystal electric field comparable at energy scales offer many emergent quantum phases~\cite{Krempa,Wan,Abhishek1,Aito,Taira}, can be achieved by tuning the strength of relative energy scales via chemical substitution~\cite{Zhu,Bikash1,Harish2,Harish3,Harish4,Vinod2,Vinod3,Vinod4,Hui2,Hui3}, reducing the particle sizes~\cite{Abhishek2,Vinod5,Lebedev,Shih,Hubert}, lattice mismatch induced strain effects~\cite{Vinod1,Chakhalian} etc. Although, theoretically $Y_2Ir_2O_7$ (YIO) is expected to be a candidate of magnetic Weyl semimetal~\cite{Wan} with all-in/all-out antiferromagnetic (AFM) ground state. Moreover, neutron diffraction and inelastic scattering measurements of YIO powder sample~\cite{Shapiro,Disseler1} does not show any sign of long-range magnetic ordering for small-moments of $Ir^{4+}(5d^5)$ within the measurement limit of instrument. However, these measurements do not rule out long-range magnetic order, but do put an upper limit for the $Ir^{4+}$ ordered moment of $\sim$0.2 $\mu_B$/Ir (for a magnetic structure with wave vector $Q \neq 0$) or $\sim$0.5 $\mu_B$/Ir (for Q = 0) based on the structural refinements of neutron powder diffraction pattern. On the other hand, muon spin relaxation ($\mu$SR) is very sensitive to probe internal magnetic fields as a result of ordered magnetic moments or random fields that are static or fluctuating (of a few Gauss) with time due to its large gyromagneic ratio.  The zero field $\mu$SR measurements of YIO powder sample shows the appearance of spontaneous muon spin precessions below transition temperature confirms the long-range magnetic ordering~\cite{Disseler1,Disseler2,Fernandez,Julia}. Recently, the glass-like state~\cite{Harish1} in YIO has also been reported, where the chemical doping of magnetic $Ru^{4+}(4d^4)$ ion~\cite{Harish2}, non-magnetic $Ti^{4+}(3d^0)$ ion~\cite{Harish2} at magnetic $Ir^{4+}(5d^5)$-site and the substitution of magnetic $Pr^{3+}(4f^2)$ for the non-magnetic $Y^{3+}(4d^0)$-site~\cite{Harish3} separately, enhances the magnetic relaxation rate. Moreover, it have also been shown that YIO exhibit weak ferromagnetic (FM) component along with a large AFM ground state~\cite{Zhu,Aito,Taira,Hui1,Hui2,Hui3}. Despite these advances, a conclusive understanding of the precise nature of magnetic state is not yet fully established. 

In this manuscript, we attempt to bridge the lacking by our latest finding by gradual substitution of magnetic ion $Cr^{3+}(3d^3)$ at $Ir^{4+}(5d^5)$ magnetic site in geometrically frustrated AFM pyrochlore iridates YIO, i.e. $Y_2Ir_{2-x}Cr_xO_7$ (YICO). We find that YIO shows GP along with the cluster-glass like state and these properties enhance on increasing the doping concentration. This substitution would produce following effects: (I) It would bring magnetic impurity in the $Ir^{4+}$ sublattice, (II) It would alter the concentration of charge carrier which in turn change the $3d-5d$ magnetic exchange interaction between the local $Cr^{3+}$ moments (randomly distributed in the $Ir$-sublattice) and itinerant $Ir^{4+}$ conduction electrons, (III) Since $Cr^{3+}(3d^3)$ have relatively larger strength of electronic correlation and low value of spin-orbit coupling compared to $Ir^{4+}(5d^5)$, the substitution will likely to induce disorder. Our results show the presence of magnetically ordered rare regions at $T_C < T < T^\ast$, support the formation GP-like state in YIO sample. The strength of magnetically ordered rare regions enhances as doping concentration of Cr increases. Interestingly, the nature of interacting spins in the GP regime founds to be Ising-like. In addition, the cluster glass-like state at low temperature is also observed.

\section{Experimental Method}

Polycrystalline samples of the YICO series (x = 0.0, 0.05, 0.1, 0.2) were synthesized using the conventional solid state reaction route following the protocols described elsewhere~\cite{Vinod3,Vinod6}. High purity stoichiometric amount of starting materials $Y_2O_3$, $IrO_2$ and $Cr_2O_3$ were mixed, ground, pelletized and heated in air at 1000~$^0$C for 100~h, at 1050~$^0$C for 200~h with several intermediate grinding with heating and cooling rate 3~$^0$C/min. Room temperature powder x-ray diffraction (XRD) pattern were measured to check the structural phase formation of the samples using PANalytical XPertPRO diffractometer with Cu-K$_\alpha$ radiation ($\lambda$ = 1.54056~\AA). The x-ray photoelectron spectroscopy (XPS) measurements were recorded using a PHI 5000 Versa Probe II system with an energy resolution of 0.02~eV and step size 0.05~eV. dc magnetic measurements were performed between 2-350~K with a 14~T Quantum Design Physical Property Measurement System (PPMS) using vibrating sample magnetometry (VSM) mode. ac susceptibility measurement was carried out in a cryogenic S700X model Superconducting Quantum Interference Device (SQUID) magnetometer down to 5~K at frequencies 50~Hz and 1~kHz with an excitation field H$_{ac}$ = 0.1~Oe and dc drive field H$_{dc}$ = 1000~Oe.  
 
\section{Results and Discussion}

\begin{figure}
	\centering
	\includegraphics[width=\linewidth]{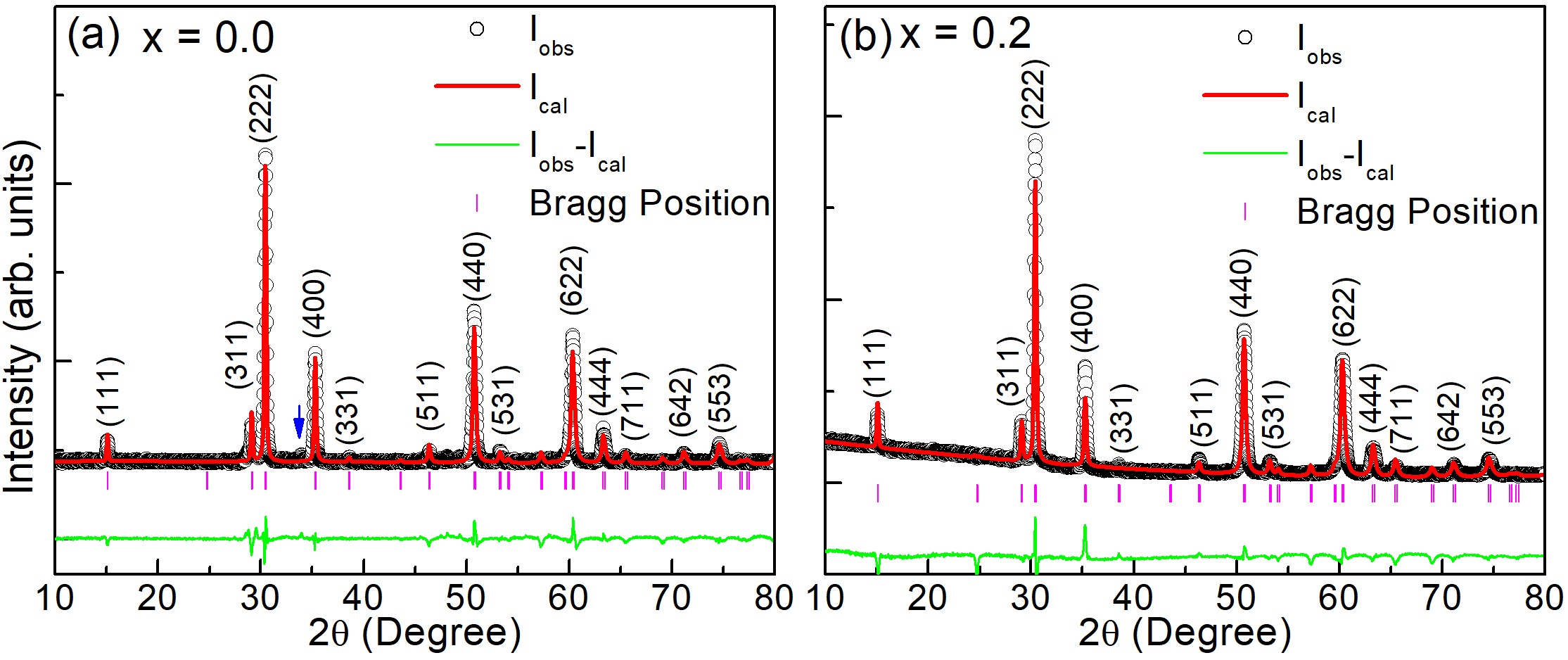}\\
	\caption{Room temperature XRD pattern of (a) x = 0.0, and (b) x = 0.2 samples. Arrow represents $Y_2O_3$ parasitic phase. }\label{fig:xrd}
\end{figure}

\begin{figure}
	\centering
	\includegraphics[width=\linewidth]{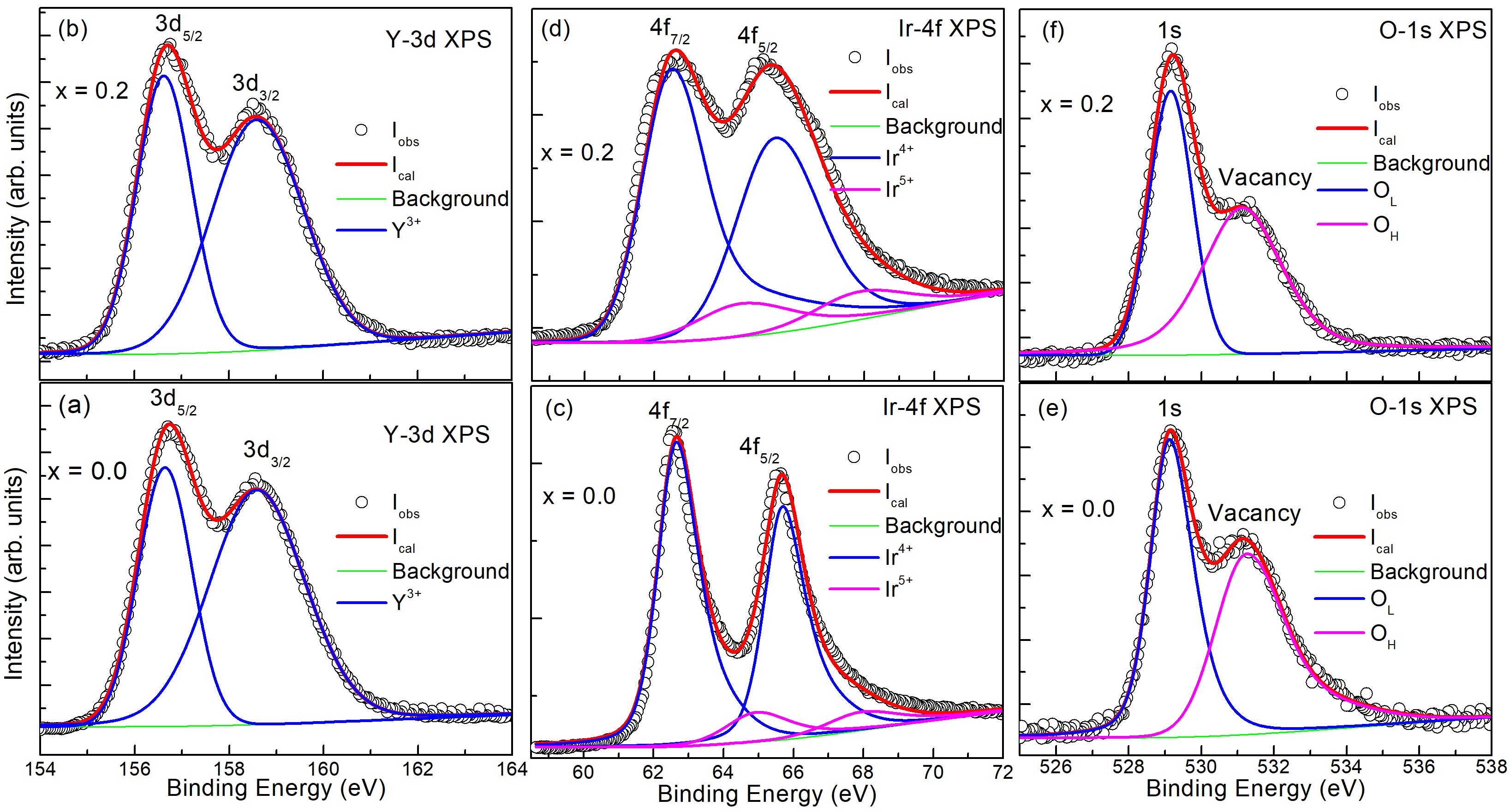}\\
	\caption{Deconvoluted XPS spectra measured at room temperature of (a) Y-3d, (b) Ir-4f, and (c) O-1s of two representative samples x = 0.0 and 0.2 of the series YICO.}\label{fig:xps}
\end{figure}

The powder XRD patterns measured at room temperature of two representative samples x = 0.0 and 0.2 of YICO series are shown in Figs.~\ref{fig:xrd}a-b. The XRD pattern was analyzed by Rietveld refinement using the software FULLPROF, confirming the phase purity of the samples. Refinement shows a pyrochlore cubic crystal structure with Fd$\bar{3}$m space group. For x = 0.0 sample, a small impurity phase of $Y_2O_3$ was detected in the sample shown by arrow in Fig.~\ref{fig:xrd}a which are in agreement with earlier literatures~\cite{Zhu,Hui1}. For x = 0.2 sample, impurity phase was not found. $Y_2O_3$ is diamagnetic in nature, hence it would not affect magnetic properties of the YICO series. The lattice parameter a, $Ir-O-Ir$ bond angle obtained from refinement for x = 0.0 and 0.2 samples are 10.16~\AA, 120$^0$ and 10.00~\AA, 108$^0$, respectively. It suggests reduction of lattice constant and bond angle with substitution of Cr. Figure~\ref{fig:xps}a-f display the XPS spectra of x = 0.0 and 0.2 samples. Y-3d XPS peaks shown in Fig.~\ref{fig:xps}a-b show a single feature suggest the presence of Y$^{3+}$ in both samples which is in agreement with earlier reports~\cite{Vinod2,Vinod3,Vinod4}. Figure~\ref{fig:xps}c-d show deconvoluted Ir-4f XPS spectra which can be fitted with two sets of doublets related to the contribution from Ir$^{4+}$ and Ir$^{5+}$ which is in agreement with reports~\cite{Vinod2,Vinod3,Vinod4,Abhishek2,Vinod5}. We find that for x = 0.2 sample, the contribution from Ir$^{5+}$ is increased, indicating co-presence of mixed oxidation states i.e., $Ir^{4+}$ and $Ir^{5+}$. Further support of mixed valence sates can also be seen in O-1s XPS spectra [Fig.~\ref{fig:xps}e-f]. The peaks centered at binding energies 529~eV and 531.1~eV are associated as lower O$_L$ and higher O$_H$ binding energy peaks, respectively~\cite{Bikash1,Vinod4,Vinod5}. The O-1s XPS spectra can be fitted with a set of doublet. We observed that the ratio of O$_H$ and O$_L$ is enhanced in x = 0.2 sample support the presence of mixed oxidation states of metal ion. 

\begin{figure}
	\centering
	\includegraphics[width=\linewidth]{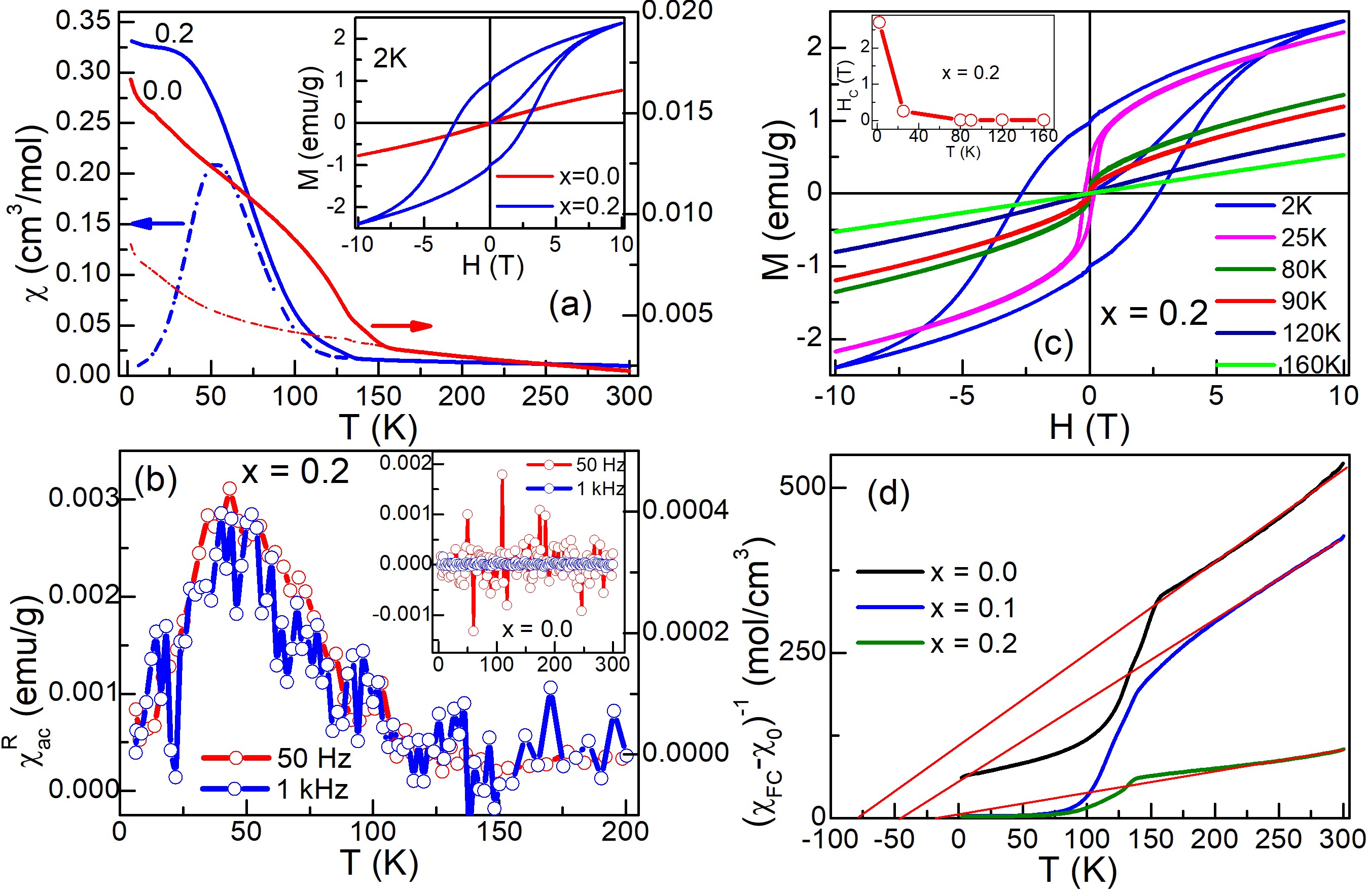}\\
	\caption{(a) Variation of magnetic susceptibility ($\chi = M/H$) with temperature measured in the presence of H = 1~kOe following the zero-field-cooled (ZFC) and field-cooled (FC) protocol shown by short dash dot line and solid continuous line, respectively; inset shows magnetization as a function of magnetic field. (b) Real part of the ac susceptibility as a function of temperature measured at 50~Hz (right y-axis) and 1~kHz (left y-axis) for x = 0.2 sample; for x = 0.0, it is shown in inset, (c) Magnetic hysteresis loops of x = 0.2 sample measured at various temperatures; inset represents temperature dependent coercive field ($H_C$), (d) $(\chi_{FC} - \chi_0)^{-1}$ as a function of temperature; red solid continuous lines are a guide to the eyes.}\label{fig:mt}
\end{figure}

Temperature dependent magnetic susceptibilty of two representative samples x = 0.0, and x = 0.2 of YICO series is shown in Fig.~\ref{fig:mt}a. For the x = 0.2 sample, the ZFC magnetic susceptibility $\chi_{ZFC}$ increases gradually to a maximum around peak temperature $T_f$ and then decreases monotonically as temperature reduces. A clear bifurcation between $\chi_{FC}$ and $\chi_{ZFC}$ curves at an irreversibility temperature $T_{irr}$ is observed. The cusp in $\chi_{ZFC}$ vs T curve is more prominent for x = 0.2 compound ($\sim$ 50~K) as compared to x = 0.0 ($\sim$ 130~K, hardly visible in $\chi_{ZFC}$ vs T curve). The value of $T_{irr}$ and $T_f$ is given in Table~\ref{tab:mueffective}. A sharp rise in $\chi_{FC}$ and $\chi_{ZFC}$ curves below $T_{irr}$ for both the samples suggest formation of FM clusters. Further, ferromagnetic transition temperature $T_C$ can be estimated using the minima in $\chi_{ZFC}$ vs T curves, while maximum in $\chi_{ZFC}$ vs T curve indicates antiferromagnetic transition $T_N$. The determination of transition temperature is done following the protocol reported by the same author~\cite{Vinod3}. The estimated values of $T_C$ is given in Table~\ref{tab:mueffective}. Figure~\ref{fig:mt}b shows real component of ac-susceptibility as a function of temperature for sample x = 0.2, suggests a well defined cusp at temperature $\sim$ 50~K similar to freezing temperature T$_f$ of $\chi_{ZFC}$ vs T curve. However, the quality of data is not good enough to fix any frequency dependence characterizing conventional spin glass-like feature. ac-susceptibility of x = 0.0 sample is shown in inset of Fig.~\ref{fig:mt}b but the signal was very weak. Figure~\ref{fig:mt}c shows the M vs H curves measured at several temperatures of x = 0.2 sample. The estimated value of $H_C$ decreases  monotonously as temperature increase [inset of Fig.~\ref{fig:mt}c] and disappear above $T_C \sim 70~K$, however, it shows huge enhancement of $H_C$ in the glassy state. It suggests evolution of glassy behaviour arises likely due to randomly arranged Cr ions in Ir-sublattice. Eventually, it is interesting to see that $T_{irr}$ is far away from the $T_f$, and hysteresis in M vs H curves below $T_C$ suggest formation of cluster glass (CG)-like behaviour~\cite{Yong} possibly due to the coexistence of mixed oxidation states of Ir. As a result, FM-PM transition arises due to FM super- and double exchange interactions via Ir$^{4+}$-O$^{2-}$-Cr$^{3+}$ or Ir$^{4+}$-O$^{2-}$-Ir$^{5+}$ path, respectively and glassy state at $T_f$ because of Ir$^{4+}$-O$^{2-}$-Ir$^{4+}$ and Cr$^{3+}$-O$^{2-}$-Cr$^{3+}$ AFM couplings.

\begin{table}
	\centering
	\caption{Parameters obtained from the magnetization data. FM transition temperature $T_C$ is estimated from the minima in first derivative of $\chi$ vs T curve. $H_C$ estimated from M(H) curve.}\label{tab:mueffective}
	\begin{tabular}{|c|c|c|c|c|}
		\hline
		Sample &x = 0.0& x = 0.05& x = 0.1& x = 0.2\\
		\hline
		$T^\ast$(K) & 230 & 240 & 240 &245 \\
		\hline
		$T_{irr}$(K) & 160 &155 &145 &140 \\
		\hline
		$T_C$(K) & 130 & 65 & 67 &70 \\
		\hline
		$T_f$(K) &130 &46 &48 &50 \\
		\hline
		-$\theta_{CW}$(K) &75 &70 &40 &10 \\
		\hline
		$\mu_{eff}^{exp}$($\mu_B$/f.u.)& 2.27 &2.62 &2.95 &4.92 \\
		\hline 
		$\mu_{eff}^{theo}$($\mu_B$/f.u.)&1.73 &2.57 &2.68 &2.9 \\
		\hline
		$H_C$(T)&0.037 &1.66 &2.74 &2.78 \\
		\hline
		$GP_{Norm.} = \frac{T^\ast-T_C}{T_C}$ & 0.77 & 2.88 & 3.50 &3.85 \\ 
		\hline
	\end{tabular}
\end{table}

Further, the susceptibility of YICO series was analysed using the modified CW law, $\chi$ = $\chi_0$ + $\frac{C}{T-\theta_{CW}}$; where $C$, $\theta_{CW}$ and $\chi_0$ are the Curie constant, CW temperature, and temperature independent susceptibility, respectively. A fit using CW law well above $T^\ast$ shown in the Fig.~\ref{fig:mt}d give the best fitted values of important parameters are listed in Table~\ref{tab:mueffective}. The negative value of $\theta_{CW}$ for all the samples indicates antiferromagnetic correlations. The observations from a closer inspection of Fig.~\ref{fig:mt}d are :- (1) non-linearity in the $(\chi_{FC} - \chi_0)^{-1}$ vs T curve at $T_{irr} < T < T^\ast$, likely due to the presence of strong crystal field. (2) $(\chi_{FC} - \chi_0)^{-1}$ vs T follows the CW law above $T^\ast$. (3) Larger value of experimental effective magnetic moments $\mu_{eff}^{exp}$ than theoretical $\mu_{eff}^{theo}$. Theoretical effective magnetic moment can be estimated using the formula reported elsewhere~\cite{Hui2,Hui3,Vinod2}, i.e., $\mu_{eff}^{theo} = \sqrt{(2-x)\mu^2_{Ir} + x\mu^2_{Cr}}$, where $\mu^2_{Ir}$ and $\mu^2_{Cr}$ are the spin only contributed values of effective magnetic moment of Ir and Cr ions, respectively. The enhancement in $\mu_{eff}^{exp}$ as compared to $\mu_{eff}^{theo}$ suggests the formation of ferromagnetic clusters~\cite{Vinod1}. (4) A sharp downward deviation well below $T^\ast$. (5) Marginal upward deviation in $(\chi - \chi_0)^{-1}$ vs T curves from the ideal CW behaviour near $T^\ast$ just before the start of downturn with lowering of temperature, similar to other systems~\cite{Vinay,He,Saber}. These observations suggest GP-like behaviour, i.e., formation of magnetically ordered rare regions in the global PM matrix at $T_C < T < T^\ast$.
 
\begin{figure}
	\centering
	\includegraphics[width=\linewidth]{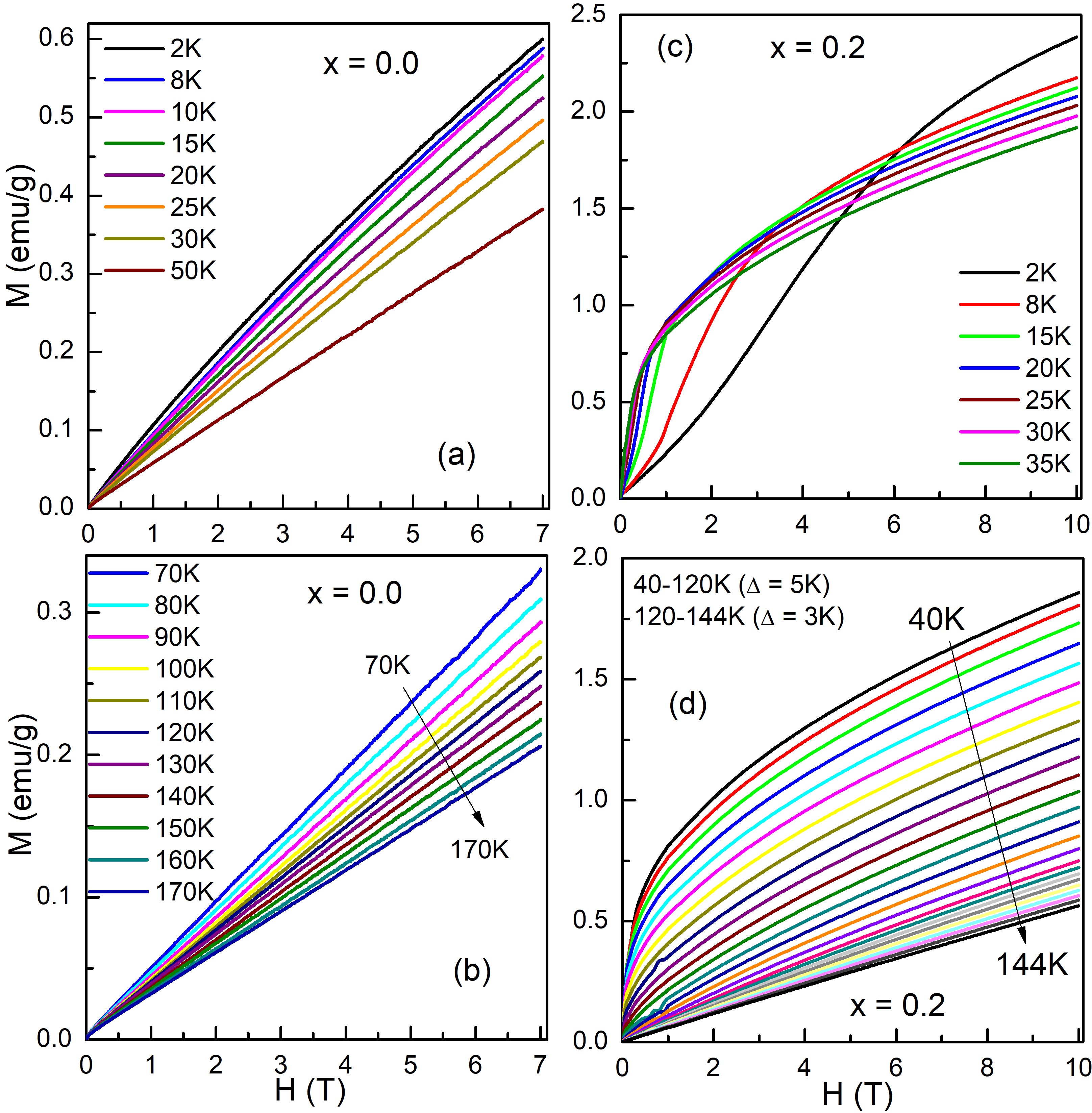}\\
	\caption{Magnetic field dependence of virgin magnetization isotherms of x = 0.0 (a), (b) and x = 0.2 (c) and (d) samples.}\label{fig:mh}
\end{figure}

\begin{figure}
	\centering
	\includegraphics[width=\linewidth]{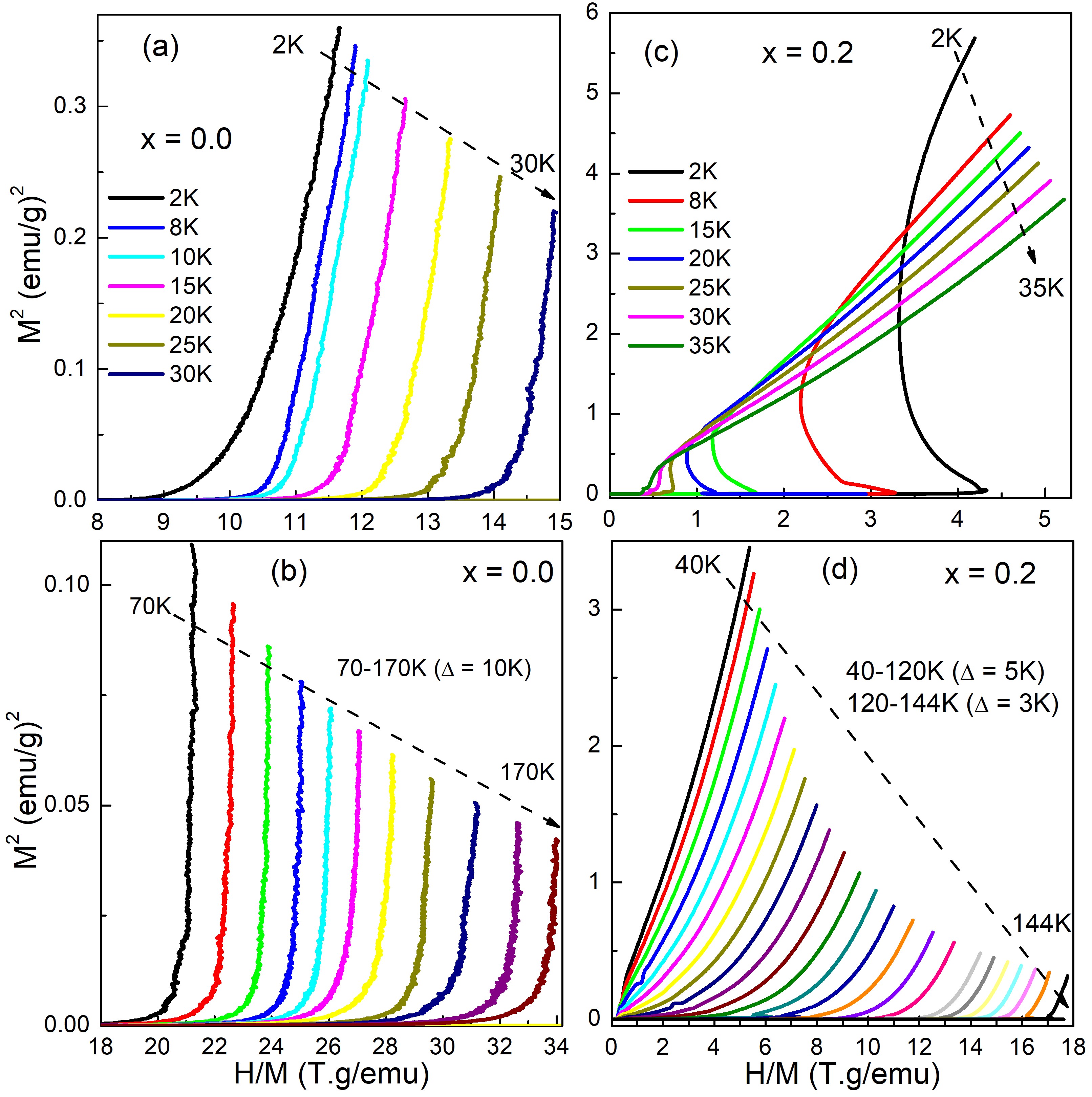}\\
	\caption{Standard Arrott plots of two representative samples x = 0.0 (a), (b), and x = 0.2 (c), (d) of YICO series.}\label{fig:arrott}
\end{figure}

Further, isothermal M vs H measurements at distinct temperatures have been carried out. All M vs H data were recorded in ZFC condition and before each successive magnetic isotherms, sample was heated above $T^\ast$ to eliminate the magnetic history. This protocol is {\it{important}} for pyrochlore iridates to measure the M vs H isothermal magnetization. Figure~\ref{fig:mh}a-d show the virgin isotherm magnetization curves for x = 0.0 (up to 7~T) and x = 0.2 (up to 10~T) samples at closely spaced temperatures. M vs H curves show monotonic enhancement with field without any sign of saturation for both the samples. The absolute values of M at 7~T and 2~K are estimated to be 0.6~emu/g (0.07~$\mu_B$) and 2.4~emu/g (0.23~$\mu_B$) for x = 0.0 and 0.2 samples, respectively. For sample x = 0.0, close inspection indicate a slight convex-like behaviour at low temperature [Fig.~\ref{fig:mh}a], while linear nature [Fig.~\ref{fig:mh}b] at high temperatures. On the other hand, sample x = 0.2 shows a crossover at low field up to 35~K [Fig.~\ref{fig:mh}c]. Surprisingly, we do not find linear behaviour in M-H curve up to 170~K [Fig.~\ref{fig:mh}d], suggest the presence of non-paramagnetic regime.

Further, to examine the strength and nature of the complex magnetic interactions, the conventional Arrott plots~\cite{Vinod4} ($M^2$ vs $H/M$) are shown in Fig.~\ref{fig:arrott}a-d. A negative intercept is observed for both the samples x = 0.0 and 0.2, suggest the absence of spontaneous magnetization. This information is important because long-range type AFM correlation possibly exhibit weak FM clusters. For x = 0.2 sample, the strength of negative intercept on $M^2$ axis decreases as compared to x = 0.0 with lowering of the temperature, suggests possibility of enhanced ferromagnetic signal. Interestingly, Arrott plot shows distorted “S”-like shape with negative curvature at low field (below 1T) shown in Fig.~\ref{fig:arrott}c, suggest the complex nature of field induced metamagnetic transition. In addition, non-straight $M^2$ curves with positive slope at low field and vertically at high field suggests the existence of short-range correlations in the proximity of AFM background in both samples.

\begin{figure}
	\centering
	\includegraphics[width=\linewidth]{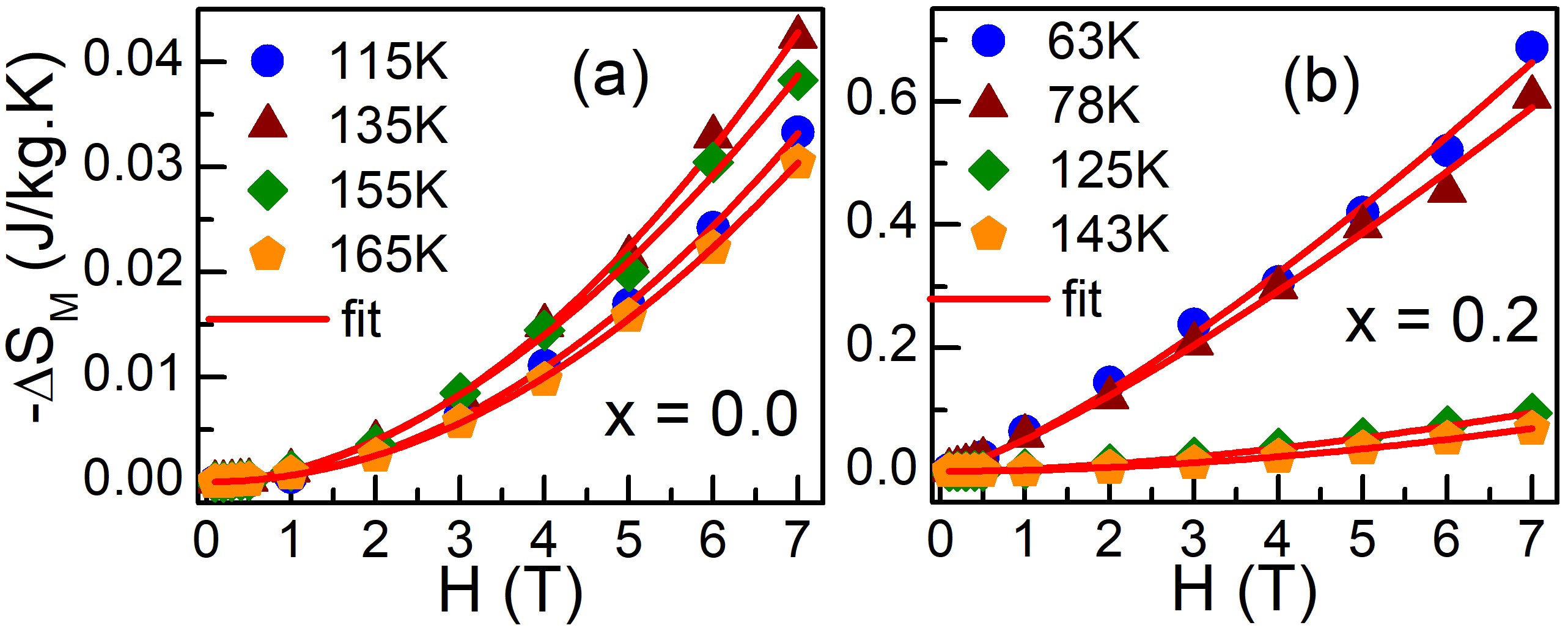}\\
	\caption{-$\Delta S_M$-H at different temperatures of (a) x = 0.0 and (b) x = 0.2 samples; red solid line represents $H^n$ dependence.}\label{fig:entppower}
\end{figure}

In order to find further support for the GP behaviour, i.e., presence of magnetic clusters in the global paramagnetic matrix ranging $T_C < T < T^\ast$, the variation in magnetocaloric entropy change $\Delta S_M = \int^H_0(\partial M/\partial T)dH$ is estimated from M vs H isotherms. It has already been shown by the same author~\cite{Vinod3} that $\Delta S_M$ vs T measured at several fields exhibit the coexistence of conventional and inverse magnetocaloric effect (MCE) in Cr substituted YIO samples, emerging due to the coexistence and competition between the ferromagnetic and antiferromagnetic clusters. It is further analysed using the prediction of mean field theory, which suggests that $\Delta S_M$ vs H should follow the power law~\cite{Ajay} behaviour $\Delta S_M \propto H^n$, where n is the local exponent of the entropy change. Further theory states that for an ideal FM system, the value of n should be 0.67 near $T_C$, $\sim$ 1 well below $T_C$ and 2 in the paramagnetic regime above $T_C$. We have fitted -$\Delta S_M$ vs H data using power law shown by red line in the Fig.~\ref{fig:entppower}a,b. The estimated values of n for the sample x = 0.0 turns out to be 1.88 (115~K), 1.92 (135~K), 1.95 (155~K), and 1.98 (165K). On the other hand for sample x = 0.2, the best fit value of n turns out to be 1.05 (63~K), 1.25 (77~K), 1.65 (125~K), and 1.85 (143~K). It is obvious that estimated value of n is less than 2 above $T_C$ for both the samples, deviates from the ideal value of n = 2 above $T_C$ for the paramagnetic regime. It suggests the presence of magnetic clusters above $T_C$ in the global paramagnetic regime in YICO series. The lower values of n for the x = 0.2 sample also suggests the higher contribution of rare region in substituted samples compared to x = 0.0.

\begin{figure}
	\centering
	\includegraphics[width=\linewidth]{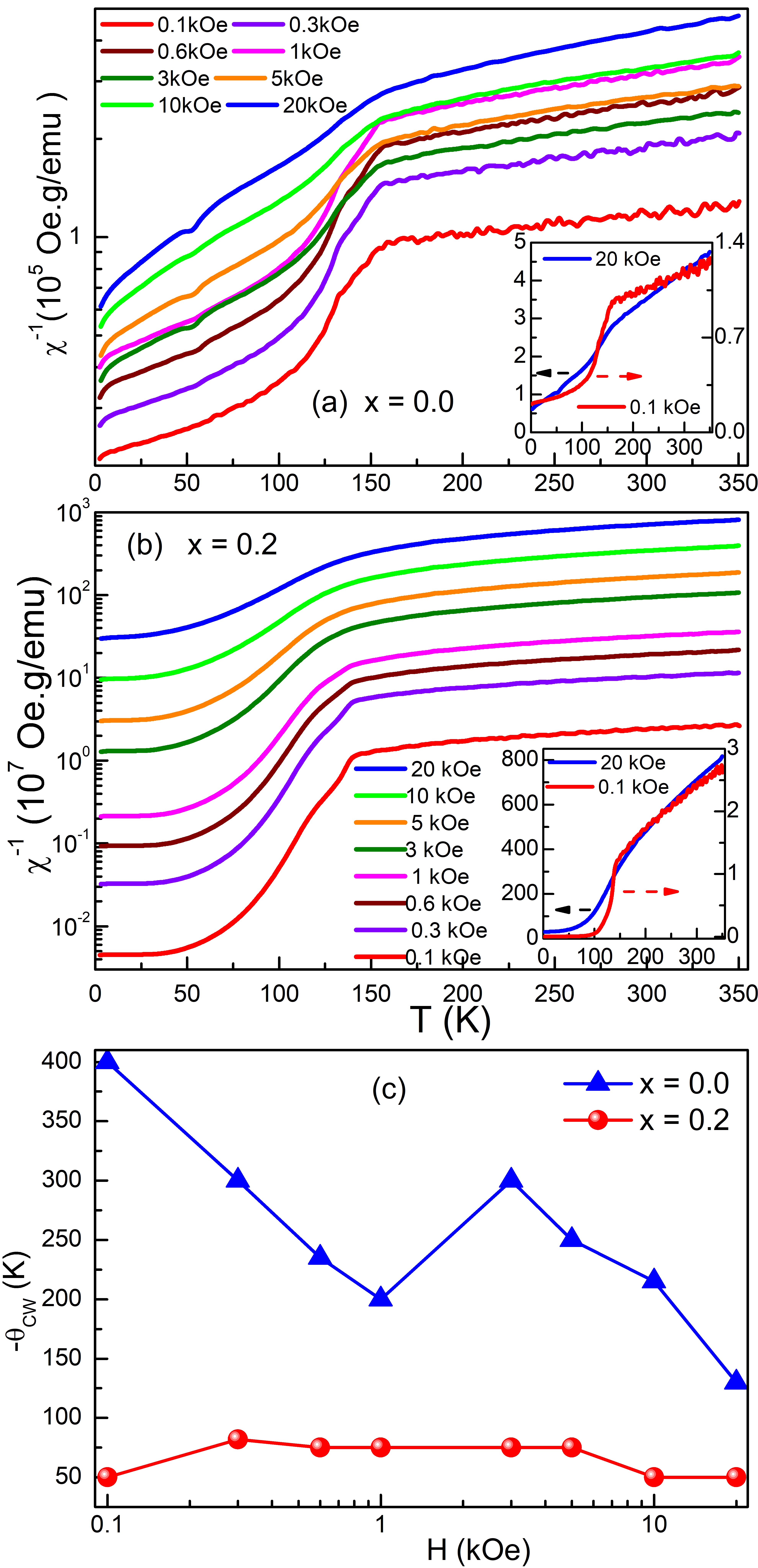}\\
	\caption{Semi-log plot of $\chi^{-1}$ vs T for the samples (a) x =0.0, (b) x = 0.2. Insets shows same at lowest and highest fields on linear scale. (c) Variation of -$\theta_{CW}$ with field.}\label{fig:gp}
\end{figure}

To elaborate the nature of probable GP, $\frac{1}{\chi}$ vs T graph were plotted at different fields. Figure~\ref{fig:gp}a, b show a pronounced field dependent sharp and sudden downward deviation on approaching the magnetic transition from the high temperature PM region. The sharpness of downturn decreases with field and suppressed at higher field; although suppression of downturn with field is not systematic for x = 0.0 sample. Such behaviour is consistent with the characteristic of GP. Generally, magnetic systems exhibiting GP show:- (I) Sharp downward deviation of $\frac{1}{\chi}$ vs T curves from conventional CW behaviour below $T^\ast$, where the sharpness of this downturn decreases as H increases. (II) Overlapping of all the $\frac{1}{\chi}$ vs T curves measured at all field in the global PM regime above $T^\ast$. It is obvious that $\frac{1}{\chi}$ vs T curves measured at all applied magnetic field do not overlap in the global PM regime above $T^\ast$. This only means that the effect of external magnetic field is beyond the simple mean-field behavior. The variation of -$\theta_{CW}$ with H for both the samples is plotted in the Fig.~\ref{fig:gp}c, indicates that the -$\theta_{CW}$ changes with field for the same sample. Moreover, it could be clusters or FM impurities, but also any other interaction that might be comparable with Zeeman energy at the given field. Thus, the GP-like state appear to be an inherent characteristic of YICO.

\begin{figure}
	\centering
	\includegraphics[width=\linewidth]{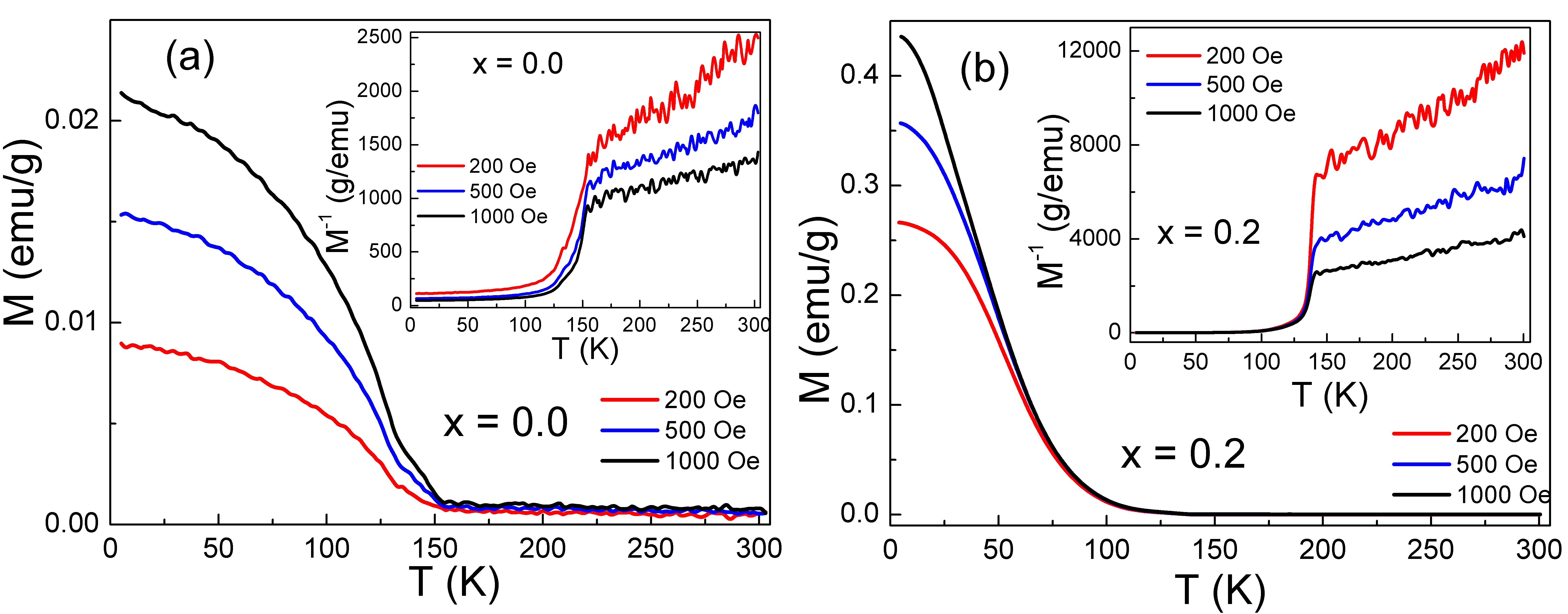}\\
	\caption{Thermoremanent magnetization (TRM) as a function of temperature  for (a) x = 0.0, and (b) x = 0.2 specimens; inset shows sharp downturn below T$^\ast$.}\label{fig:trm}
\end{figure}

Figure~\ref{fig:trm}a,b show the thermoremanent magnetization (TRM) of x = 0.0, 0.2 samples measured at different field. Sample was cooled down to low temperature in presence of applied field. After 10~sec waiting time, M vs T curve is measured in a heating cycle after sudden removal of field. Being a zero field measurement, TRM is likely to be advantageous in comparison with the traditional in-field $\chi_{dc}$ vs T measurements in identification of the GP singularity as the paramagnetic contribution to the magnetization is likely to be suppressed. Around $T_{irr}$ the TRM exhibits a sharp upturn, while a sharp and sudden downturn in $M_{TRM}^{-1}$ vs T curves [insets of Fig.~\ref{fig:trm}] indicate the formation of rare regions at $T_C < T < T^\ast$.

\begin{figure}
	\centering
	\includegraphics[width=\linewidth]{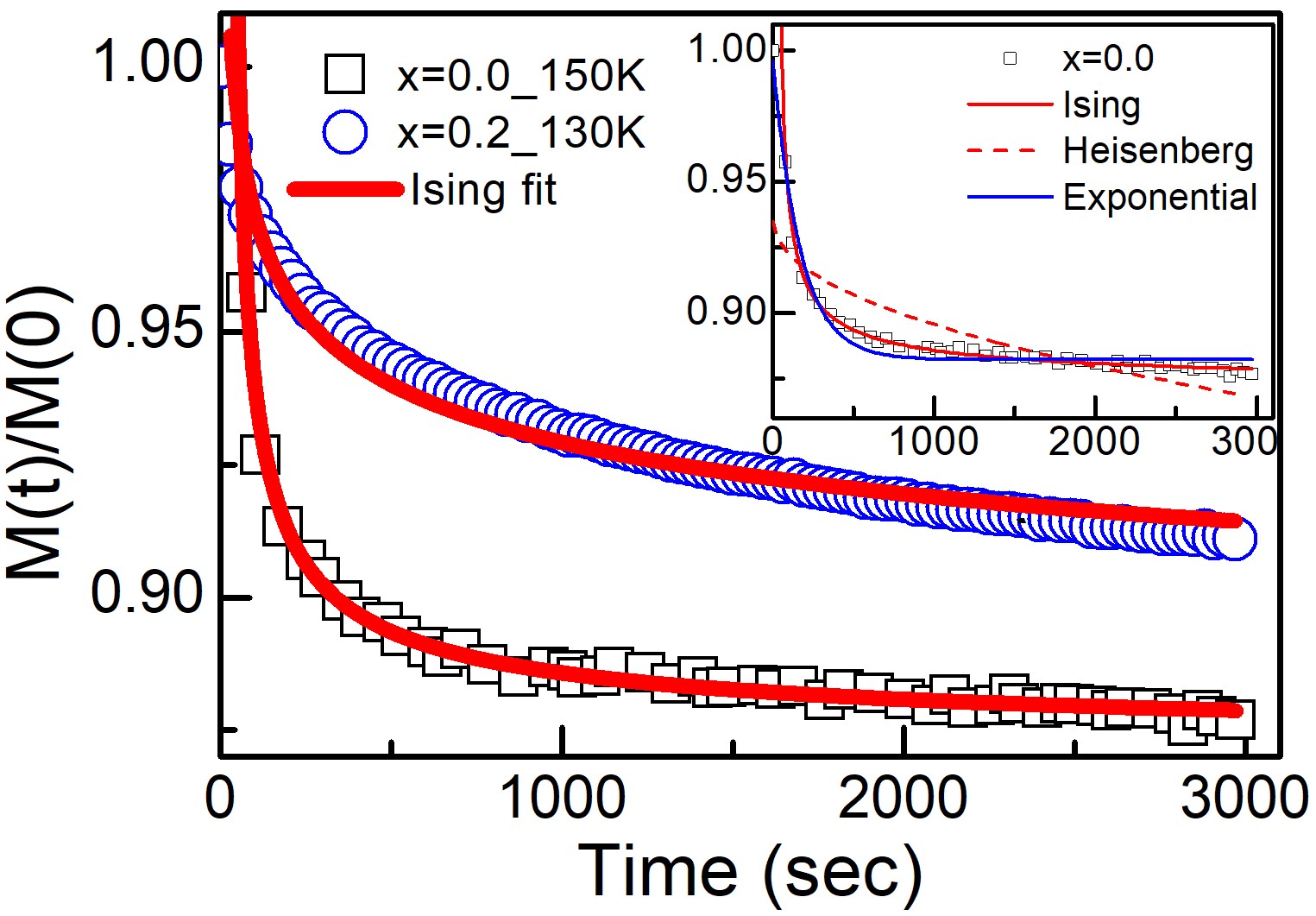}\\
	\caption{Variation of normalized IRM with time decay measured at $T_C < T < T^\ast$ for the x = 0.0 (150~K) and x = 0.2 (130~K) samples, fitted with Ising model. Inset shows fitting at 150~K of x = 0.0 sample along with their Ising, Heisenberg and exponential model.}\label{fig:relaxation}
\end{figure}

In order to find the presence of magnetically ordered rare regions, isothermal remanent magnetization (IRM) at $T_C < T < T^\ast$ is carried out. In PM regime, IRM generally fall exponentially with time, while in GP regime, IRM fall non-exponentially with time because magnetically ordered rare regions would take larger time to reverse its effective spin~\cite{Mohit,Bray}. Long back, it have been proposed that in the GP regime, spin auto-correlation function $C(t)$ has to be the form of $C(t)$ $\sim$ $exp[-A^\ast ln(t)^{d/(d-1)}]$ for Ising system and $C(t)$ $\sim$ $exp[-Bt^{1/2}]$ for Heisenberg system~\cite{Mohit,Bray}. Figure~\ref{fig:relaxation} shows the normalized IRM as a function of time for the two representative samples x = 0.0 (150~K), 0.2 (130~K). For the IRM measurement, the samples were cooled from 300~K to desired temperature in the presence of field 1~kOe. After stabilizing the temperature and waiting time up to 100~sec, variation of magnetization as a function of time (M vs. t) was measured after immediate removal of the field. Interestingly, IRM data indicate  the best fit with the Ising spin model decay scheme [Fig.~\ref{fig:relaxation}]. It shows the lack of agreement with the Heisenberg-like interactions as well as exponential decay model scheme [$M(t) = M(0) + A^\ast exp(\frac{-t}{\tau})$] as shown in the inset of Fig.~\ref{fig:relaxation}. It suggests that instead of PM state above $T_C$, system is in magnetically ordered rare region (which slows down the dynamics of spins). The observation of {\it{Ising-like}} interaction is consistent with prior reports, where the nature of spin correlation in the pyrochlore iridates was inferred to be Ising-like~\cite{Savary}.

\begin{figure}
	\centering
	\includegraphics[width=\linewidth]{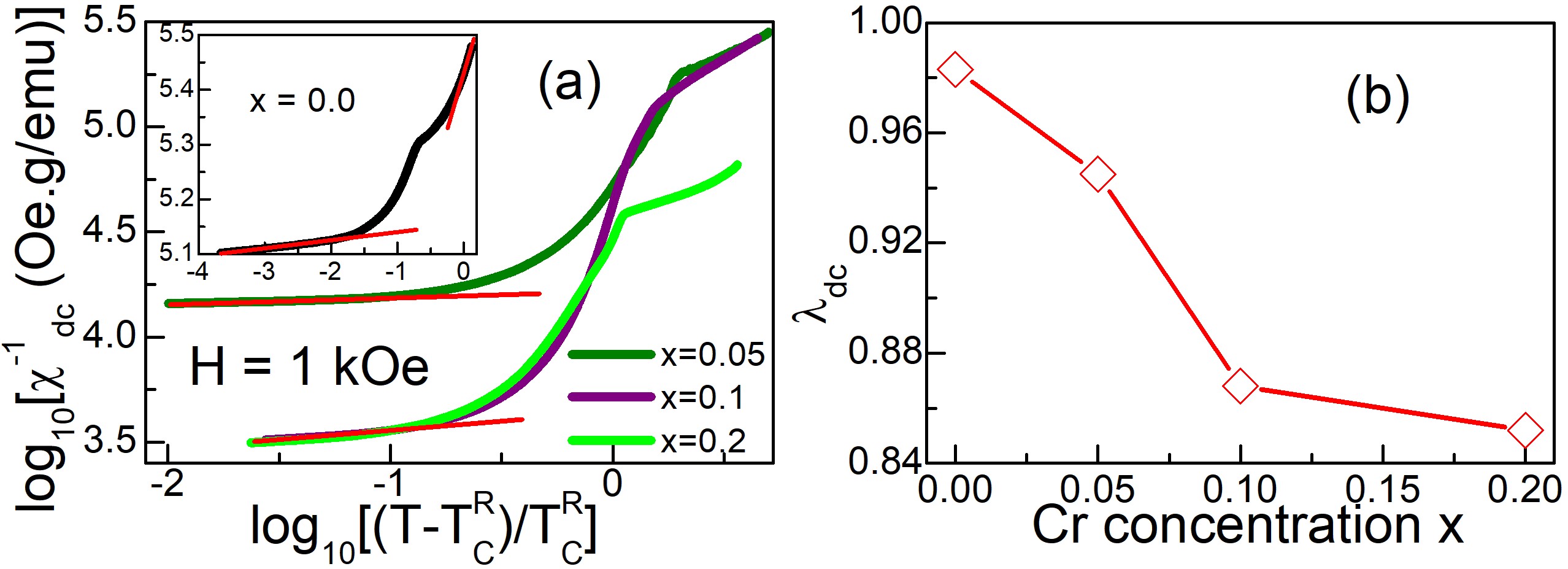}\\
	\caption{(a) The temperature dependent susceptibility data, plotted in log-log scale. For clarity, x = 0.0 sample is shown in inset. (b) Dependence of the GP parameters with doping.}\label{fig:lembda}
\end{figure}

One of the standard methods to characterize the GP behaviour is to verify the $\frac{1}{\chi}$ vs T defined by $\chi^{-1} = (T-T_C^{Rand})^{1-\lambda}$, where, the exponent $\lambda$ is positive but less than unity. Here $T_C^{Rand}$ is the critical temperature of random FM clusters lying above $T_C$ but below $T^\ast$. The exact estimation of $T_C^{Rand}$ is a serious issue. An appropriate choice should be $T_C^{Rand} = \theta_{CW}$ so that efficiently it gives rise to $\lambda \sim 0$ in the PM region~\cite{Pramanik,Jitender,Ghosh}.The value of $\theta_{CW}$ is much lower than $T_C$. So, we approximately choose $T_C^{Rand}$ = $T_C$. Figure~\ref{fig:lembda}a shows $log_{10}(\chi^{-1})$ vs $log_{10}(\frac{T-T^{Rand}_C}{T^{Rand}_C})$ plots. Fitting of the linear regime according to GP equation gives non-zero values of $\lambda$ ranging between $0 \leq \lambda \leq 1$. The value of $\lambda$ decreases systematically against $Cr$ doping concentration [Fig.~\ref{fig:lembda}b]. We further estimate the value of normalized range of GP$_{Norm}$ = $(\frac{T^\ast-T_C}{T_C})$~\cite{Pramanik,Ouyang} are given in Table~\ref{tab:mueffective}. The estimated value of GP$_{Norm}$ is greater in substituted sample (~1 order) as compared to x = 0.0 sample . Moreover, YICO series is showing significant enhancement in the value of GP$_{Norm}$ compared to other systems~\cite{Vinod1,Magen,Ouyang,Ghosh}. 

Further, we look into the possible macroscopic origin of the formation of magnetically ordered rare regions. In parent compound $Ir$-site creates local disorder in the pyrochlore crystal structure with the doping of $Cr$-ion as reported by the same authors~\cite{Vinod2,Vinod3}. As un-doped compound is a disordered system exhibiting strong distorted cubic structure, the quenched disorder (a prerequisite of the GP) is inherent in the system. There are several reports on GP, which emerged due to $B$-site~\cite{Pramanik,Arkadeb,Silva} disorder with mixed oxidation states of transition metal ions. Distortion leads to the coexistence of mixed valence states of $Ir$, i.e. $Ir^{4+}$ and $Ir^{5+}$ in YICO series. Now, the random distribution of vacancies at $Ir^{4+}$-site locally reduces the $Ir^{4+}-Ir^{4+}$ interionic bond length where short range correlations are prevalent. It leads to the coexistence of two phases within the same crystalline state consistent with other systems exhibiting geometrically frustrated AFM spin arrangement~\cite{Ouyang,Jitender,Ghosh}.

\section{CONCLUSIONS}
We investigate the experimental evidence of GP-like behaviour at $T_C < T < T^\ast$ in the $Cr$ doped geometrically frustrated antiferromagnetic pyrochlore iridates $Y_2Ir_2O_7$. In a nutshell, the important observations are as follows:- (1) Hysteresis in M vs H curve, cusp in $\chi_{ZFC}$ vs T data at lower temperature than irreversible temperature $T_{irr}$, and relaxation in M vs time curve suggest cluster-glass-like state in YICO series, (2) Larger value of $\mu_{eff}^{exp}$ than  $\mu_{eff}^{theo}$, suggests the formation GP-like FM clusters, (3) Sudden and sharp downward deviation in $\frac{1}{\chi}$ vs T curves from conventional CW behaviour below $T^\ast$, a temperature much above the long range ordering temperature $T_C$, suggests formation magnetically ordered clusters, (4) The $\frac{1}{\chi}$ vs T curves measured at all applied magnetic field do not coincide on to a single curve above $T^\ast$ in the true PM regime, (5) The lesser value of power exponent n than ideal value of 2 (for paramagnetic regime) at $T_C < T < T^\ast$ in -$\Delta S_M$ vs H curve, suggests presence of magnetic clusters, and (6) Slow dynamics of spin in IRM data at $T_C < T < T^\ast$ exhibiting Ising like interaction, favours the presence of magnetically ordered rare regions. Interestingly, all experimental observations suggests the presence of magnetically ordered rare regions in global paramagnetic matrix ranging $T_C < T < T^\ast$, favours GP-like state in YIO. The strength of magnetically ordered rare regions enhances with the substitution of Cr. Rare regions are distributed over a wide temperature regime. It is the first study of the GP along with cluster-glass-like states in the geometrically frustrated antiferromagnetic pyrochlore oxide family. The current finding is important and justify its importance among the rare materials which exhibit GP along with glass like states in AFM ordered systems, which is very limited till date.

\section{Acknowledgments}
V. K. Dwivedi thanks Prof. Soumik Mukhopadhyay (IIT Kanpur) for his guidance during this project.

\end{document}